\documentclass[12pt]{article}

\pdfoutput=1 

\usepackage[english]{babel}
\usepackage[utf8]{inputenc}
\usepackage[T1]{fontenc}

\usepackage[a4paper,top=3cm,bottom=2cm,left=3cm,right=3cm,marginparwidth=1.75cm]{geometry}

\usepackage{graphicx,epstopdf,amssymb,amsfonts,amsmath,amsthm,array,
mathrsfs,amscd}
\usepackage{subcaption}
\usepackage[colorinlistoftodos]{todonotes}
\numberwithin{equation}{section}
\usepackage[square,comma,numbers,sort&compress]{natbib}
\usepackage{float}
\usepackage
{hyperref}

\newcommand{\pbs}[1]{\let\temp=\\#1\let\\=\temp}
\numberwithin{equation}{section}
\def\be{\begin{equation}}\def\ee{\end{equation}}
\def\cvp{\raise 2pt\hbox{,}} 
 \def\tr{\mathop{\rm tr}\nolimits}

 \def\d{{\rm d}} 
 \def\uN{\text{U}(N)}

\def\la{\lambda}
\def\oD{\text{O}(D)}
\def\gB{\mathcal B}

\def\npb#1#2#3{{\it Nucl.\ Phys.\ }{\bf B #1} (#2) #3}

\def\prl#1#2#3{{\it Phys.\ Rev.\ Lett.\ }{\bf #1} (#2) #3}
\def\jhep#1#2#3{{\it J. High Energy Phys.\ }{\bf #1} (#2) #3}
\def\prd#1#2#3{{\it Phys.\ Rev.\ }{\bf D #1} (#2) #3}
\def\prb#1#2#3{{\it Phys.\ Rev.\ }{\bf B #1} (#2) #3}

\def\jmp#1#2#3{{\it J.\ Math.\ Phys.\ }{\bf #1} (#2) #3}

\def\imath#1#2#3{{\it Invent math }{\bf #1} (#2) #3}

\def\ahp#1#2#3{{\it Ann.\ Henri Poincar\'e }{\bf #1} (#2) #3}

\begin{document}

{\pagestyle{empty}
\parskip 0in
\

\vfill
\begin{center}

{\LARGE More on the New Large $D$ Limit of Matrix Models}

%

\vspace{0.4in}

Tatsuo A{\scshape zeyanagi},${}^1$ \,Frank F{\scshape errari},${}^{1,2}$ \,Paolo G{\scshape regori},${}^{1,3}$ \\ 
L\ae titia L{\scshape educ}${}^{1}$  and \,Guillaume V{\scshape alette}${}^{1}$ 
\\

\medskip
${}^1${\it Service de Physique Th\'eorique et Math\'ematique\\
Universit\'e Libre de Bruxelles (ULB) and International Solvay Institutes\\
Campus de la Plaine, CP 231, B-1050 Bruxelles, Belgique}

\smallskip

${}^{2}${\it Fields, Gravity and Strings\\
Center for the Theoretical Physics of the Universe\\
Institute for Basic Sciences, Daejeon, 34047 South Korea}

\smallskip

${}^{3}${\it Universit\`a di Torino, Dipartimento di Fisica \\
and I. N. F. N. - sezione di Torino \\
Via P. Giuria 1, I-10125 Torino, Italy}


\smallskip
{\tt tatsuo.azeyanagi@ulb.ac.be,  frank.ferrari@ulb.ac.be, \\paolo.gregori@ulb.ac.be, 
laetitia.leduc@ulb.ac.be, \\guillaume.valette@ulb.ac.be}
\end{center}
\vfill\noindent

In this paper, we extend the recent analysis of the new large $D$ limit of matrix models to the cases where the action contains arbitrary multi-trace interaction terms as well as to arbitrary correlation functions. We discuss both the cases of complex and Hermitian matrices, with $\text{U}(N)^{2}\times\text{O}(D)$ and $\text{U}(N)\times\text{O}(D)$ symmetries respectively. In the latter case, the new large $D$ limit is consistent for planar diagrams; at higher genera, it crucially requires the tracelessness condition. For similar reasons, the large $N$ limit of tensor models with reduced symmetries is typically inconsistent already at leading order without the tracelessness condition. We also further discuss some interesting properties of purely bosonic models pointed out recently and explain that the standard argument predicting a non-trivial IR behaviour in fermionic models \`a la SYK does not work for bosonic models. Finally, we explain that the new large $D$ scaling is consistent with linearly realized supersymmetry. 

\vfill

\medskip
%
\begin{flushleft}
\today
\end{flushleft}
\newpage\pagestyle{plain}
\baselineskip 16pt
\setcounter{footnote}{0}

}

\section{\label{introduction} Introduction and summary}

In a series of recent developments, interesting toy models for quantum black holes have been built and studied. The first class of models is based on large $N$ fermionic systems with quenched disorder and was proposed by Kitaev \cite{Kitaev}, building on previous studies in the condensed matter literature by Sachdev, Ye and others \cite{SachdevYeetal}. The second class of models is based on large $N$ tensor theories and were first proposed by Witten in \cite{witten}, building on the tensor model technology developed by Gurau and collaborators \cite{Gurauetal}. There is a rapidly growing literature on this subject, see e.g.\ \cite{SYKmore, SYKsusy, wittenmore}. These models are able to capture very non-trivial properties of black holes, including the quasi-normal behavior and chaos \cite{miscrefsBH}. The advantage of the tensor models over models with quenched disorder is that they are genuine quantum theories at finite $N$; in particular, there is no need to limit the investigations to self-averaging quantities.

Both models with quenched disorder and tensors remain, however, rather exotic. String theory, via the open/closed string duality, singles out unambiguously matrix models in the 't~Hooft's large $N$ limit as being the favored candidates to describe quantum black holes. Matrix models are ubiquitous in string theory simply because the two indices of the matrices are the Chan-Paton factors associated with the two end points of open strings. It is very difficult to find a similar interpretation for tensors of rank three or higher.

The models originating from D-brane constructions always involve several bosonic matrices $X_{\mu}$, $1\leq\mu\leq D$, which describe motion transverse to the brane worldvolume. The index $\mu$ naturally transforms in the fundamental representation of $\text{O}(D)$, which is the rotation group in the directions orthogonal to the branes. The full symmetry is usually $\uN\times\oD$, the $\uN$ part being gauged. These models must be studied in the planar $N\rightarrow\infty$ limit and superficially seem to be much more difficult to solve than models with quenched disorder or tensors. 

Recently, it was shown in \cite{Fer1} that the above-mentioned large $N$, $\text{O}(D)$-invariant matrix models have a new large $D$ limit which is both analytically tractable and captures the essential physics associated with the sum over planar diagrams. The limit is ``new'' in the sense that it does not coincide with the well-known large $D$ limit of $\text{O}(D)$-invariant vector models because, crucially, the large $D$ scaling of some coupling constants is enhanced.\footnote{For the 
use of the standard large $D$ limit with no enhancement in the context of matrix models, see for example~\cite{oldlargeD}.} 
This implies that many more Feynman diagrams contribute at large $D$ than what one would find in a vector model and the result yields the expected continuous spectrum of states and chaotic behaviour. As explained in \cite{Fer1}, the new large $D$ limit could also be related to the large space-time dimension limit of general relativity studied in \cite{emparan}.

The consistency of the new large $D$ limit is ensured by remarkable and unexpected constraints the genus of a Feynman diagram puts on the highest possible power of $D$ the diagram can be proportional to. The technology involved to prove some of these results is directly imported from the tensor model literature \cite{CarrozzaTanasa}, which may not be surprising since our matrices are objects with three indices $X^{a}_{\mu\, b}$. However, there are important differences, both conceptual and technical, with the tensor models. The fact that the matrix indices $a,b$ on the one hand and $\mu$ on the other hand transform with respect to different groups is conceptually fundamental, since the group associated with the matrix indices must always be gauged in string theory. Moreover, the large $D$ expansion does not coincide with the large $N$ expansion of tensor models, because it is made at fixed genus. In particular, the large $N$ and large $D$ limits do not commute, the large $N$ limit must always be taken first. 

The purpose of the present note is to complement the analysis of \cite{Fer1}, both at the technical level and on the possible applications of the models. First, we generalize the discussion to arbitrary multi-trace interaction terms and to arbitrary multiply-connected interaction bubbles.\footnote{See below for definitions.} Multi-trace terms have been shown to be important in holographic contexts \cite{multitraceAdSCFT}, but the analogue of this useful generalization does not seem to have been studied before in the context of tensor models. The results we obtain also play a role in the new large $N$ and large $D$ limits for general matrix-tensor models studied in \cite{FRV}.\footnote{The results of the present note relevant for \cite{FRV} were obtained before the development of the general theory presented in \cite{FRV}.}  We also discuss the general structure of the large $N$ and large $D$ expansions of arbitrary correlation functions. We emphasize the special features of models with reduced symmetry $\uN\times\oD$ instead of $\uN^{2}\times\oD$. The large $D$ limit remains well-defined for planar diagrams. However, without any further constraint on the matrices, it is inconsistent at higher genera; similarly, models involving symmetrized tensors, 
proposed recently in the literature \cite{klebanov, klebanov2}, do not have a consistent large $N$ limit. Interestingly, when the tracelessness condition is added on the matrices and/or the tensors, the basic obstructions to the existence of the limits are waived. Finally, we emphasize that our results apply to a very wide and interesting class of matrix theories in space-time dimensions $0\leq d\leq 3$. In particular, the new large $D$ scaling of coupling constants is consistent with ordinary linearly realized supersymmetry. We also explain some crucial differences between the well-studied fermionic models \`a la SYK and bosonic models, giving more details on some properties first pointed out in \cite{AFS}.

\section{\label{subsection2-1} Definition of the models}

Our models are $\oD$-invariant matrix theories. The basic variables are complex or Hermitian matrices $X_{\mu}$ transforming in the fundamental representation of $\oD$. When we deal explicitly with complex matrices, we always assume that the models are also invariant under a $\text{U}(N)_{\text L}\times\text{U}(N)_{\text R}$ symmetry acting as $X_{\mu}\mapsto U_{\text L}X_{\mu}U_{\text R}^{-1}$. In the purely Hermitian case, this symmetry is reduced down to a single $\uN$ factor and the matrices transform in the adjoint representation. The matrices $X_{\mu}$ may carry additional ``flavor'' labels, may be bosonic or fermionic and may live in various number of space-time dimensions. This additional information is irrelevant for our purposes.

For many applications, it is important to gauge the $\text{U}(N)$ symmetries of the models, whereas the $\text{O}(D)$ symmetry is ungauged. The explicit gauging can be straightforwardly performed and does not change our discussion in any non-trivial way, so we shall not mention it any further in the following. Note that, in the leading large $N$ and large $D$ approximations, the gauging is altogether irrelevant.

Our results can be straightforwardly generalized to other types of symmetries and matrix ensembles. For example, a special case of our analysis corresponds to models invariant under $\text{U}(D)$ instead of $\text{O}(D)$; similar methods can be applied to models of real matrices with orthogonal or symplectic gauge symmetries and $\text{O}(D)$ vector symmetry, etc.\ A completely general formalism is described in \cite{FRV}.

The Lagrangian of the models are of the form
\be\label{lagrangian}
L = ND\Bigl(\text{Kinetic Term} - \sum_{a}N^{1-t(\gB_{a})}\tau_{a}I_{\gB_{a}}(X) \Bigr)\ .\ee
The kinetic term is $\tr X_{\mu}\mathscr{D} X_{\mu}^{\dagger}$ for some wave operator $\mathscr D$ that does not act on the $\text{U}(N)$ or $\text{O}(D)$ indices. The $I_{\gB_{a}}(X)$ are $\text{O}(D)$ invariant $t(\gB_{a})$-trace interaction terms, labeled by $\gB_{a}$, with associated 't~Hooft's coupling constants $\tau_{a}$. They can be written as
\be\label{interterm}
I_{\gB_{a}}(X) =\prod_{i=1}^{t(\gB_{a})}\tr\bigl(X_{\mu_{1,i}} X_{\mu_{2,i}}^\dagger \cdots X_{\mu_{2r_{i}-1,i}} X_{\mu_{2r_{i},i}}^{\dagger}\bigr)\, ,\ee
where the $O(D)$ indices are contracted pairwise and summed over. In particular, the degree of an interaction vertex, which is the number of matrices $X$ and $X^{\dagger}$ entering in \eqref{interterm}, is always even. Note that the models studied in \cite{Fer1} correspond to single-trace interactions $t(\gB_{a})=1$.

\section{\label{verandgraSec} Vertices and graphs}

\begin{figure}
\centering
\includegraphics[width=6in]{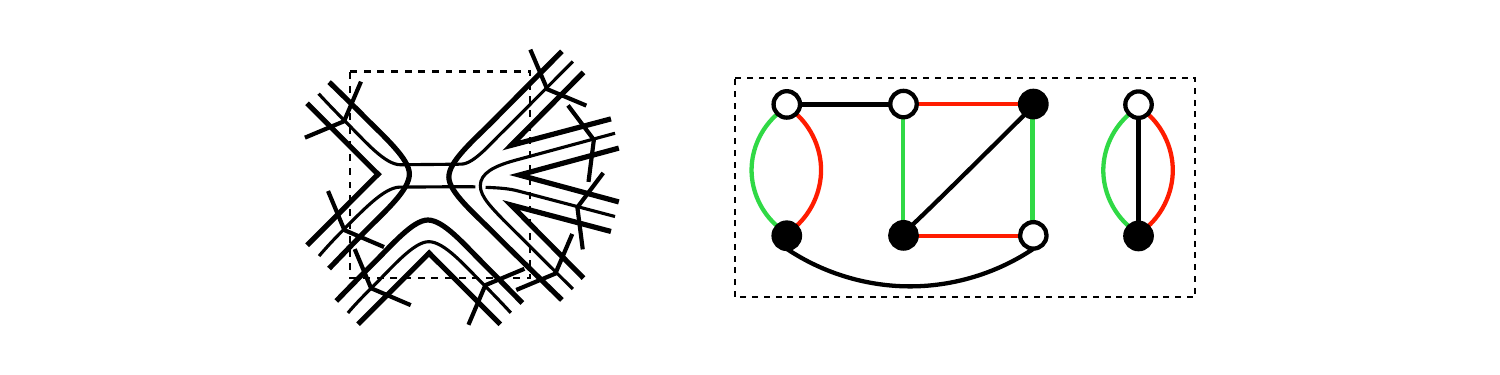}
\caption{Fat graph and colored graph for the interaction vertex $\tr X_{\mu}X_\nu^\dagger \,\tr X_\mu X_\rho^\dagger X_\nu X_\rho^\dagger\,\tr X_\sigma X_\sigma^\dagger$, with $c=2$, $t=3$ and $g=1/2$.}
\label{figure1}
\end{figure}

As in \cite{Fer1}, we use two graphical representations for each interaction vertex $\gB_{a}$: the standard stranded fat graph representation and the three-colored bubble graph representation. We shall often denote by $\gB_{a}$ either the interaction term itself or the associated three-colored graph. Our detailed conventions are exactly the same as in \cite{Fer1} and we shall not repeat them here. Simply note that the colors (green, red, black) are also denoted by $(1,2,3)$. To any interaction vertex $\gB_{a}$, we assign: the number of connected components $c(\gB_{a})$ of the bubble; the number of traces $t(\gB_{a})$, which is also the number of $(12)$-faces of the associated bubble, $t(\gB_{a})=F_{12}(\gB_{a})$ and the genus $g(\gB_{a})$ of the interaction, given by Euler's formula in terms of the total number of faces $F(\gB_{a}) = F_{12}(\gB_{a})+F_{13}(\gB_{a})+F_{23} (\gB_{a})$ and vertices $V(\gB_{a})$ of the bubble, $2c(\gB_{a})-2g(\gB_{a})=F(\gB_{a})-\frac{1}{2}V(\gB_{a})$. A typical interaction term is depicted in Fig.~\ \ref{figure1}. When an interaction vertex has several connected components, as it is the case in the figure, we insert it in a dashed rectangular box to emphasize the fact that it represents a unique Feynman diagram vertex.

Similar to the interaction vertices, the Feynman diagrams can also be represented either by a stranded graph or by a four-colored graph. The colored Feynman graphs are obtained by representing the propagators as lines of a new color, say violet (or 0). In the case of complex matrix models, the fat graph propagators are oriented, say from $X^{\dagger}$ to $X$ and this implies that the violet lines of the colored graph respect the bipartite structure of the graph (i.e.\ they join vertices of different types). In the case of Hermitian matrices, the fat graph propagators are no longer oriented and thus in general, the violet lines of the colored graph do not respect the bipartite structure. However, as explained in \cite{Fer1}, if the fat graph is planar, then it is always possible to assume that they do respect the bipartite structure.\footnote{The argument in \cite{Fer1} applies without change to the more general case of multi-trace interactions considered here and thus will not be repeated.} This is a crucial property that allows one to extend the results obtained in the complex case to the planar Hermitian case.

\section{\label{largeNDSec} The large $N$ and large $D$ limits}

To define the large $N$ and large $D$ limits of our models, we introduce new couplings $\la_{a}$, related to the couplings $\tau_a$ appearing in the Lagrangian \eqref{lagrangian} by
\begin{equation}
\label{scaling}
\tau_{a}= D^{t(\gB_{a})-c(\gB_{a})+g(\gB_a)}\lambda_{a} \,  ,
\end{equation}
and we decide to keep $\la_{a}$ fixed. The large $N$ limit defined this way is the usual 't~Hooft's limit, suitably generalized to the case of multi-trace interactions; note in particular that both $\tau_{a}$ and $\la_{a}$ are fixed at large $N$ but finite $D$. The large $D$ limit has the enhancement factor $D^{g(\gB_{a})}$ with respect to the standard vector model large $D$ scaling, as in \cite{Fer1}, together with a new additional factor $D^{t(\gB_{a})-c(\gB_{a})}$ that takes into account both the multi-trace structure and the fact that the interaction bubbles may be disconnected. In this section, we show that the free energy has well-defined large $N$ and large $D$ limits with the scalings \eqref{scaling}. This result extends to correlation functions, whose study is postponed to Section \ref{CorrFunSec}.

Let us consider an arbitrary vacuum Feynman diagram. We denote by $p$, $v$, $f$ and $\varphi$ the number of propagators, vertices, $\uN$ and $\oD$ faces, respectively. With the Lagrangian \eqref{lagrangian} and the scaling \eqref{scaling}, the amplitude of the diagram is proportional to
\be\label{FeynDN} N^{-p+2v-\sum_a t(\gB_a)+f}D^{-p+v+\sum_a (t(\gB_{a})-c(\gB_{a})+g(\gB_a))+\varphi} = N^{2-h} D^{1+\frac{h}{2}-\frac{\ell}{2}} \ ,\ee
where we introduced the parameters $h$ and $\ell$ defined by
\begin{align}
\label{hexp1}
h & =2+p-2v+\sum_a t(\gB_a)-f \ , \\
\label{ellexp1}
\frac{\ell}{2} & = 2+\frac{3}{2}p-2v-\frac{1}{2}\sum_a t(\gB_a) +\sum_a c(\gB_a)-\sum_a g(\gB_a)-\frac{1}{2}f-\varphi \ .
\end{align}
\subsection{\label{NpowerSec} Counting the power of $N$}

To study the power of $N$ of a given Feynman diagram $\gB$, we consider the matrix model fat graph obtained by removing the $\oD$ lines in the stranded representation. In the colored representation, it amounts to studying the three-bubble $\gB^{(3)}$ obtained by removing the edges of color 3. Since we deal with multi-trace interactions, the resulting fat graph may be disconnected. Indeed, each multi-trace vertex effectively leads to $t(\gB_a)$ single-trace vertices in the fat graph. We denote by $\tilde v$ the total number of effective single-trace vertices so that $\tilde v = \sum_a t(\gB_a)$. Besides, the number of connected components of the fat graph is the same as the number of connected components $B^{(3)}$ of the three-bubble $\gB^{(3)}$. The genus $g$ of the fat graph\footnote{We always implicitly define the genus of a multiply-connected graph as the sum of the genus of each connected component.} is then given by the usual Euler's formula
\be\label{Eulerfatgraph} 2B^{(3)}-2g=-p+f+\tilde v=-p+f+\sum_a t(\gB_a) \ . \ee
We can obtain a similar relation by studying the corresponding three-bubble $\gB^{(3)}$, whose genus is given by the relation
\be\label{Eulercolgraph}
\begin{split}
2B^{(3)}-2g(\gB^{(3)}) & = -\frac{1}{2} V(\gB^{(3)}) + F(\gB^{(3)}) \\ &
= -\frac{1}{2}V(\gB) +  F_{01}(\gB) + F_{02}(\gB)+F_{12}(\gB) \ .
\end{split}
\ee
By using the following identities that connect the quantities characterizing $\gB$ in the stranded and colored representations,
\be \label{fatcolid} 2p=V(\gB) \ ,  \quad f = F_{01}(\gB) + F_{02}(\gB) \ , \quad \sum_a t(\gB_{a}) = F_{12}(\gB) \ , \ee
it is straightforward to check that the genera of the fat graph and the colored graph coincide, 
\be \label{genusrel} g = g(\gB^{(3)}) \ . \ee

Using the above formulas, we can rewrite $h$ in \eqref{hexp1} as
\be \label{hexp2} \frac{h}{2} = g+\sum_a \bigl(t(\gB_a)-1\bigr)-B^{(3)} +1\ , \ee
which importantly shows that $h$ is non-negative since it is given by the sum of two non-negative terms,
\be \label{posterms} g\geq 0 \ , \quad 1+\sum_a \bigl(t(\gB_a)-1\bigr)-B^{(3)} \geq 0 \ . \ee
The second inequality comes from the fact that each $t(\gB_a)$-trace interaction vertex can increase the number of connected components of the fat graph by $t(\gB_a)-1$ at most. The non-negativity of $h$ ensures that the large $N$ limit \`a la 't~Hooft of models with multi-trace interactions is well-defined. For single-trace interactions, $t(\gB_a)=1$ and $B^{(3)}=1$ so that $h=2g$ as usual.

\subsection{\label{DpowerSec} Counting the power of $D$}

By generalizing the proof for the single-trace models found in \cite{Fer1}, we want to express $\ell$ given in \eqref{ellexp1} as the sum of non-negative terms. The Euler's formula \eqref{Eulercolgraph} for the three-bubble $\gB^{(3)}$ generalizes straightforwardly to the three other three-bubbles $\gB^{(0)}$, $\gB^{(1)}$ and $\gB^{(2)}$. We write them in a unified way as
\be\label{Eulercolgraph2} 2B^{(i)}-2g(\gB^{(i)}) = -\frac{1}{2} V(\gB) + \sum_{\substack{j<k \\ j,k\neq i}}F_{jk}(\gB) \ , \ee
where $i=\{0,1,2,3\}$ and $B^{(i)}$ is the number of connected components of the three-bubble $\gB^{(i)}$. By summing these equations for $i=0,1,2$ and using the following identities that complement the ones in \eqref{fatcolid},
\be \label{fatcolid2}  \varphi = F_{03}(\gB) \ , \quad \sum_a c(\gB_{a}) = B^{(0)} \ , \quad \sum_a g(\gB_a) = g(\gB^{(0)}) \ , \ee
one obtains the following relation
\begin{multline} \label{sumEuler}
g(\gB^{(1)})+ g(\gB^{(2)}) + \bigl({B}^{(01)} - {B}^{(1)} - {B}^{(0)} + B\bigr) 
+ \bigl({B}^{(02)} - {B}^{(2)} - {B}^{(0)} + B\bigr) \\
 = 2B + \frac{3}{2}p - \frac{1}{2}\sum_a t(\gB_a) - \sum_a c(\gB_a) - \sum_a g(\gB_a) -\frac{1}{2}f-\varphi \, ,
\end{multline}
where $B$ in the number of effective connected components of $\gB$ and $B^{(01)}=F_{23}(\gB)$, $B^{(02)}=F_{13}(\gB)$. Comparing \eqref{sumEuler} with \eqref{ellexp1}, we get
\begin{multline} \label{ellexp2}
\frac{\ell}{2} = g(\gB^{(1)})+ g(\gB^{(2)}) + \bigl({B}^{(01)} - {B}^{(1)} - {B}^{(0)} + B\bigr) 
+ \bigl({B}^{(02)} - {B}^{(2)} - {B}^{(0)} + B\bigr)\\
 + 2\Bigl(1 + \sum_a \bigl(c(\gB_a)-1\bigr) - B \Bigr)\, .
\end{multline}
The first two terms on the right hand side are manifestly non-negative. The third and fourth terms are also non-negative using the connectivity inequality
\be \label{ineq}B^{(ij)}-B^{(i)}-B^{(j)} +B\geq 0\, , \ee
which is proven in \cite{Fer1} for the case $B=1$, the case $B>0$ being a straightforward generalization (see also the discussion in \cite{FRV}). Finally, the last term is also non-negative,
\be \label{ineq2} 1+\sum_a \bigl(c(\gB_a)-1\bigr)-B \geq 0 \ , \ee
because each interaction vertex with $c(\gB_a)$ connected components increases the number of effective connected components of $\gB$ by $c(\gB_a)-1$ at most (Eq.\ \eqref{ineq2} can also be viewed as a consequence of a connectivity inequality of the form \eqref{ineq}, see \cite{FRV}).

In conclusion, we have shown that $\ell$ is a non-negative integer. This proves that the large $D$ expansion is well-defined at any fixed power of $N$. In the case of single-trace interactions only, we have that $c(\gB_a)=1$ and $B=1$ so that the expression \eqref{ellexp2} for $\ell$ matches the one found in \cite{Fer1}.

\subsection{\label{expLOgraphSect} Form of the expansions and leading order graphs}

The large $N$ expansion of the free energy reads
\be \label{Fexp} F = \sum_{h\in\mathbb N}F_h N^{2-h} \ , \ee
where the $F_h$ are $N$-independent coefficients. Each $F_h$ is itself expanded at large $D$ in powers of $1/\sqrt{D}$ as
\be \label{Fhexp} F_h = \sum_{\ell\in\mathbb N}F_{h,\ell} D^{1+\frac{h}{2}-\frac{\ell}{2}} \ , \ee
with $D$-independent coefficients $F_{h,\ell}$. In particular, we see that the highest possible power of $D$ for diagrams of given $h$ is bounded above by $1+h/2$. It is this crucial property that makes the limit exist. However, if diagrams of arbitrary $h$ are considered, there is no such upper bound. This implies that the limit $N\rightarrow\infty$ must always be taken first and then the $D \rightarrow \infty$ limit next, at each order in the $1/N$ expansion. The non-commutativity of the two limits is a central property of the new large $D$ limit introduced in \cite{Fer1}.

\begin{figure}
\centerline{\includegraphics[width=6in]{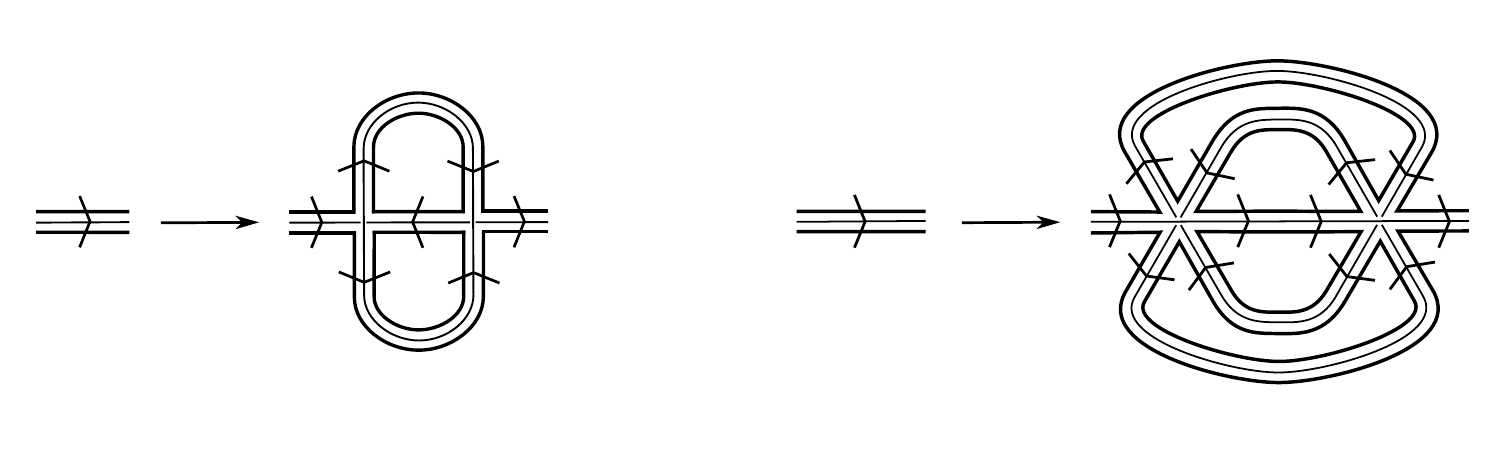}}
\caption{Melonic moves for the $\tr X_{\mu}X_{\nu}^{\dagger} X_{\mu}X_{\nu}^{\dagger}$ and $\tr X_{\mu}X_{\nu}^{\dagger}X_{\rho} X_{\mu}^{\dagger}X_{\nu}X_{\rho}^{\dagger}$ interactions.}\label{figure2}
\end{figure}

Leading order graphs, called generalized melons in \cite{FRV}, must have $h=0$ and $\ell=0$. These conditions require in particular the planarity of the three-colored graphs $\gB^{(1)}$, $\gB^{(2)}$ and $\gB^{(3)}$. Typically, they can be built by applying an arbitrary number of so-called melonic moves, which amounts to replacing internal lines by a more complicated structure, starting from the one-loop ring vacuum graph. Examples of melonic moves for the interactions $\tr X_{\mu}X_{\nu}^{\dagger} X_{\mu}X_{\nu}^{\dagger}$ and $\tr X_{\mu}X_{\nu}^{\dagger}X_{\rho} X_{\mu}^{\dagger}X_{\nu}X_{\rho}^{\dagger}$ are depicted on Fig.\ \ref{figure2}. We let the reader check explicitly that these moves do not change the powers of $N$ and $D$ (i.e.\ the values of $h$ and $\ell$).

\begin{figure}
\centerline{\includegraphics[width=6in]{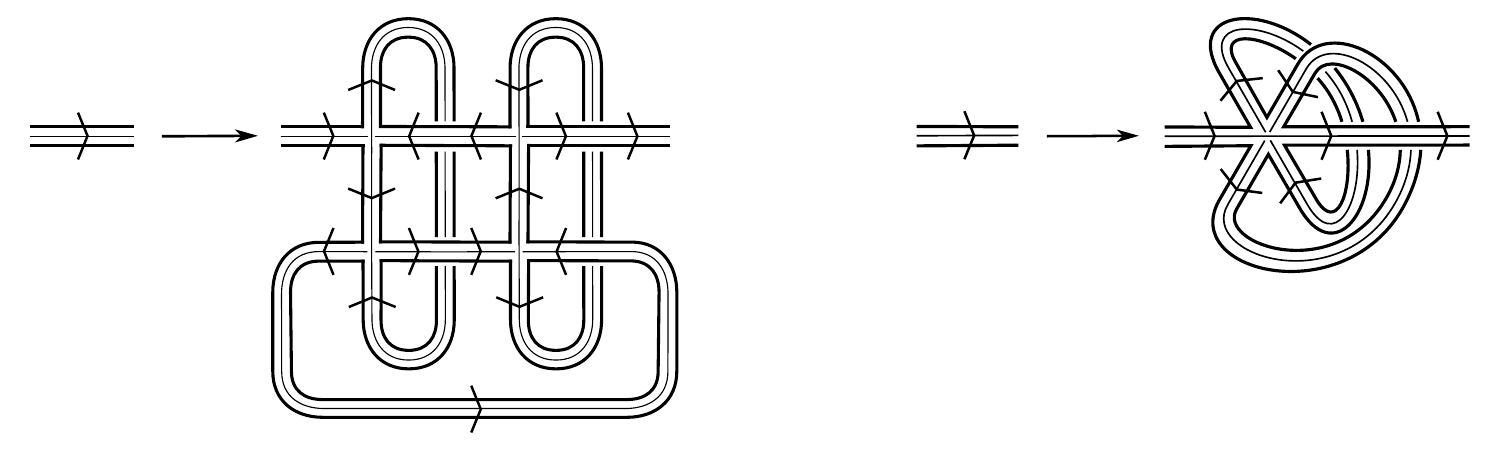}}
\caption{Moves increasing the genus by one unit at fixed $\ell$ for the $\tr X_{\mu}X_{\nu}^{\dagger} X_{\mu}X_{\nu}^{\dagger}$ and $\tr X_{\mu}X_{\nu}^{\dagger}X_{\rho} X_{\mu}^{\dagger}X_{\nu}X_{\rho}^{\dagger}$ interactions.}
\label{figure3}
\end{figure}

Leading order graphs at fixed genus $g>0$, on the other hand, are planar only with respect to the colors 0, 1, 3 and 0, 2, 3. For single-trace interactions they are proportional to $D^{1+g}$. Families of leading graphs at fixed genus can be obtained, for example, using the moves depicted on Fig.\ \ref{figure3} an arbitrary number of times. It is easy to check that these moves increase the genus by one unit but leave $\ell$ unchanged. One can of course also use the moves of Fig.\ \ref{figure2} to generate more leading graphs at fixed $g$.

\begin{figure}
\centerline{\includegraphics[width=6in]{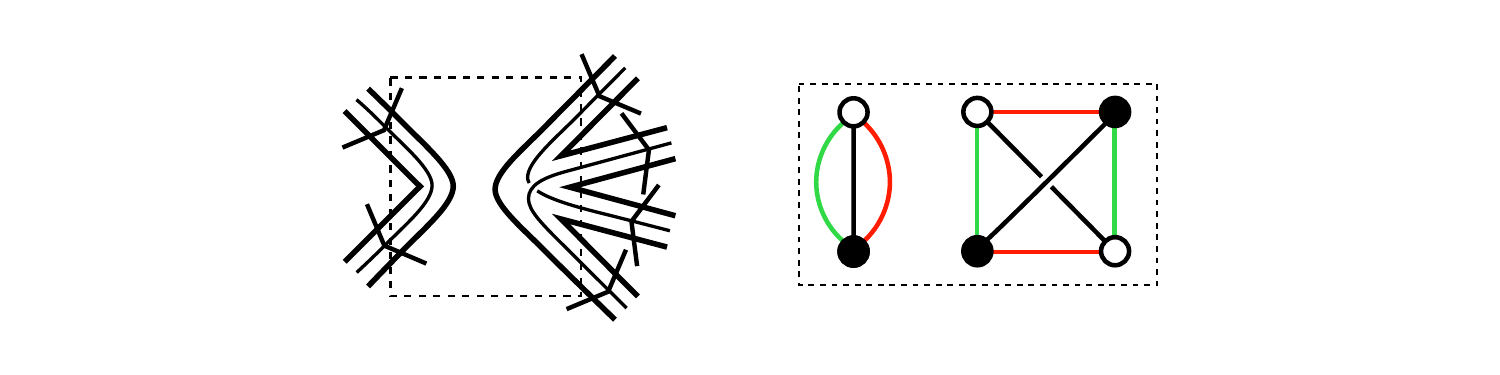}}
\caption{Stranded graph and colored graph for the interaction vertex $\tr X_{\rho}X_{\rho}^{\dagger}\,\tr X_{\mu} X_{\nu}^{\dagger} X_{\mu} X_{\nu}^{\dagger}$ discussed in the main text, with $c=2$, $t=2$ and $g=1/2$.}
\label{figure5-ex}
\end{figure}

\begin{figure}[ht!]
\centerline{\includegraphics[width=6in]{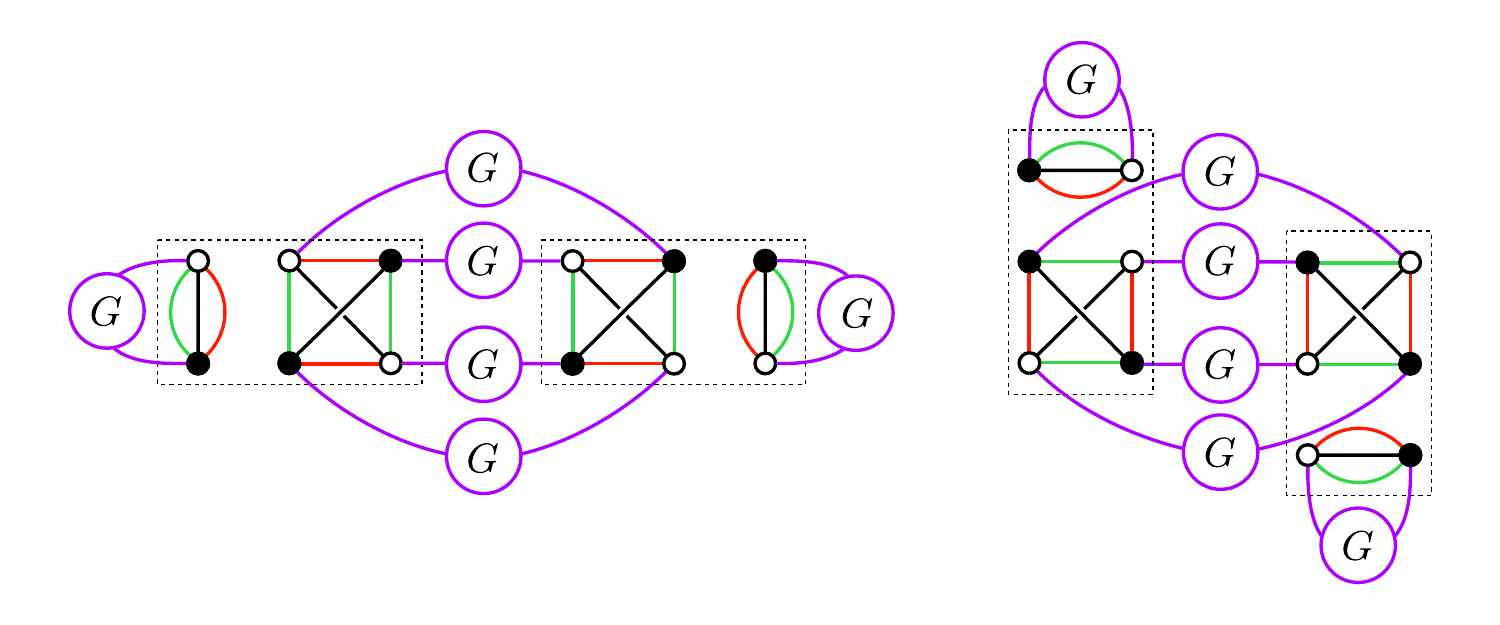}}
\caption{Structure of the leading order vacuum graphs for the multi-trace model with interaction vertex $\tr X_{\rho}X_{\rho}^{\dagger}\,\tr X_{\mu} X_{\nu}^{\dagger} X_{\mu} X_{\nu}^{\dagger}$. $G$ is the leading two-point function.}
\label{figure5-LO}
\end{figure}

\begin{figure}[ht!]
\centerline{\includegraphics[width=6in]{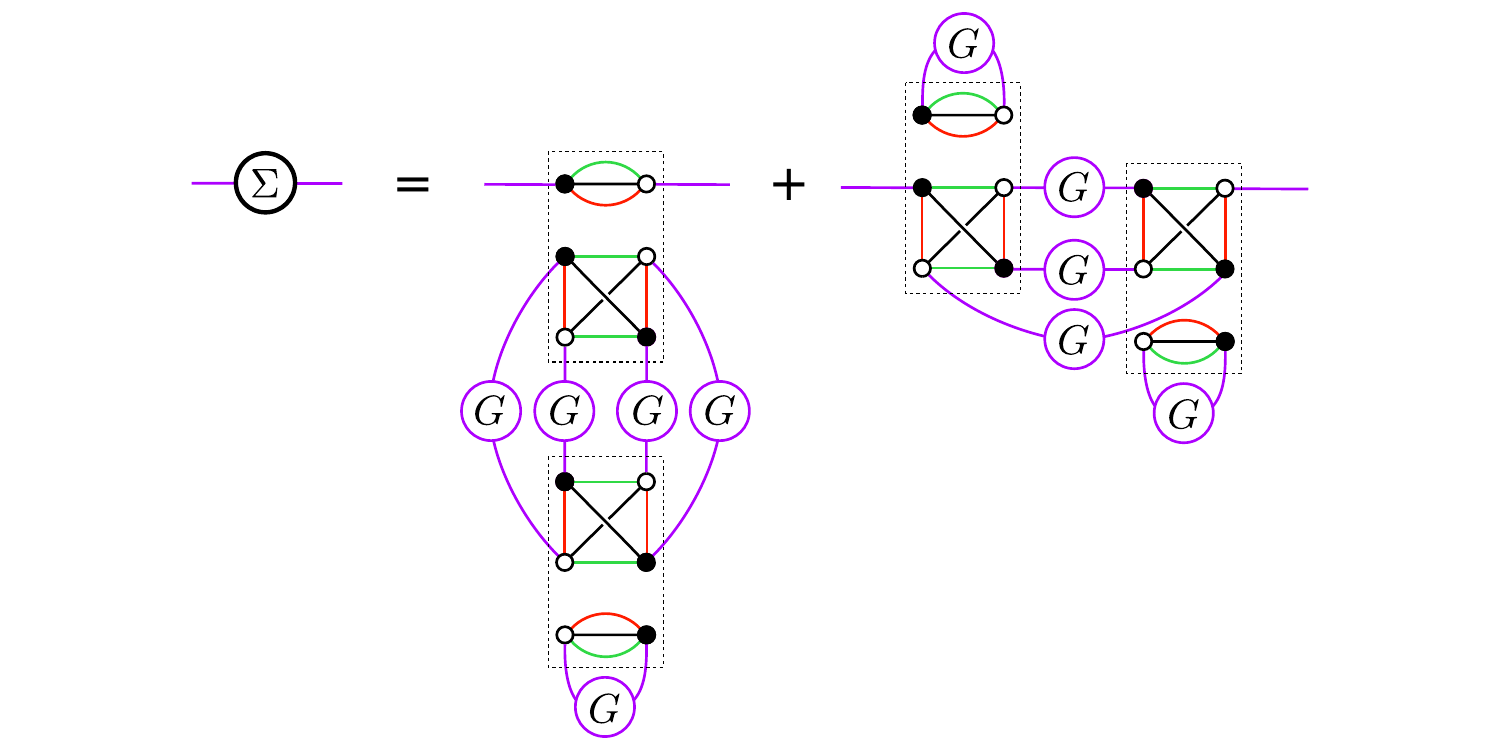}}
\caption{Diagrammatic representation of the Schwinger-Dyson equation \eqref{SDEq}.}
\label{figure5-SDEq}
\end{figure}

To illustrate the case of a multi-trace interaction, consider the interaction vertex depicted in Fig.~\ref{figure5-ex}. The leading order graphs $h=\ell=0$ must be maximally disconnected and each connected component must be a leading order graph for the model involving the effective single-trace interactions $\tr X_{\rho}X_{\rho}^{\dagger}$ and $\tr X_{\mu} X_{\nu}^{\dagger} X_{\mu} X_{\nu}^{\dagger}$. We can then straightforwardly apply the results of \cite{CarrozzaTanasa}. The leading vacuum graphs have the structure depicted in Fig.\ \ref{figure5-LO}, the two-point function being determined by the Schwinger-Dyson equation with the self-energy given by Fig.\ \ref{figure5-SDEq} or, equivalently, in terms of the equation (in the quantum mechanical case)
\begin{multline} \label{SDEq}
\Sigma(t-t')  =  (-1)^\sigma 2 \lambda^2 \delta(t-t') \int dt_1 \, G^2(t_1)G^2(-t_1) \, G(0) \\
 + (-1)^\sigma 4 \lambda^2 \, G^2(t-t')G(-t+t') \, G^2(0)\, .
\end{multline}
In this equation, $\Sigma$ is the self-energy and $\sigma=0$ or 1 depending on whether we are dealing with complex fermions or bosons.

\subsection{\label{RedSymSec} On cases with reduced symmetry}

Up to now, we have focused on models with $\uN^{2}\times\oD$ symmetry. The fact that a symmetry group is associated with each individual index is crucial in the standard tensor model technology, for which $D=N$. An interesting question is whether non-trivial and consistent large $N$ and large $D$ limits exist in models with reduced symmetry.

The first rigorous argument showing that this is the case was given in \cite{Fer1}, where the example of Hermitian matrices was studied. For Hermitian matrices, the symmetry group is reduced from $\uN^{2}\times\oD$ down to $\uN\times\oD$ and all the usual tools based on colored graphs superficially seem to break down. But the argument in \cite{Fer1}, which generalizes straightforwardly to the multi-trace interactions considered in the present paper, showed that the large $D$ limit is still well-defined for the sum over \emph{planar} diagrams. The same argument would also work for real symmetric or antisymmetric matrices, again at the planar level.

\begin{figure}
\centering
\includegraphics[width=4in]{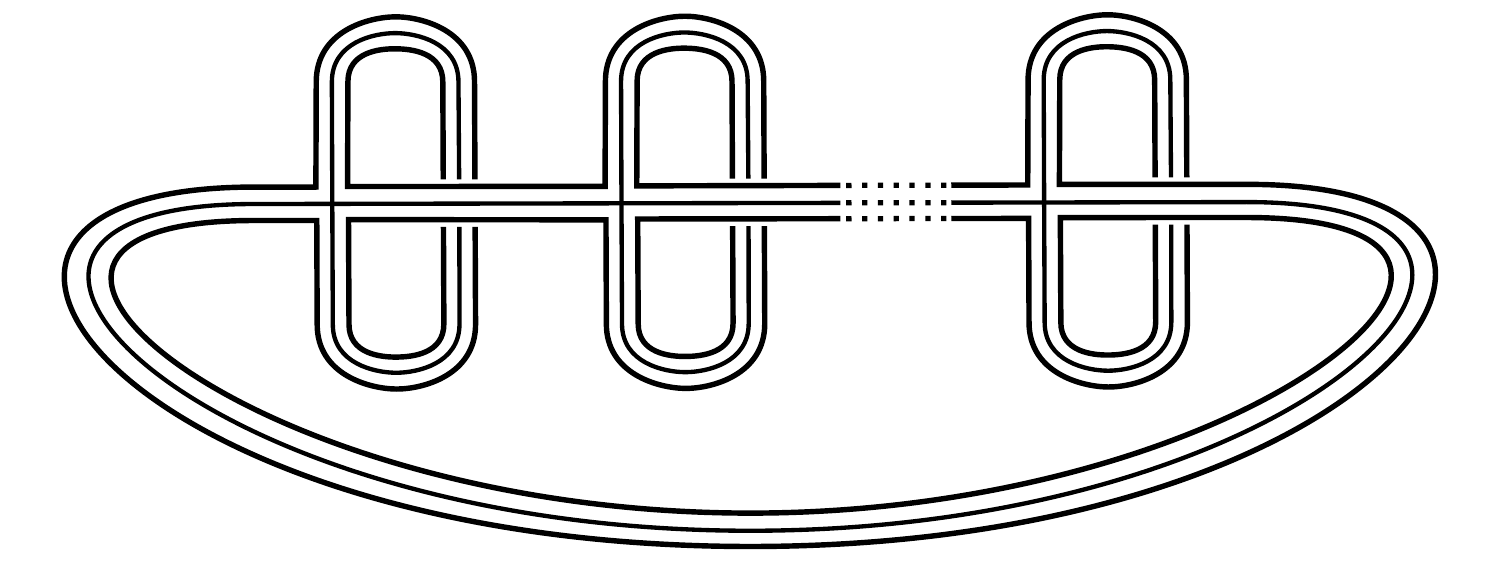}
\caption{Example of graphs at genus one in the Hermitian models (or real symmetric models) that are proportional to an arbitrarily high power of $D$.}
\label{figure6}
\end{figure}

However, it is not difficult to prove that the large $D$ limit for general Hermitian matrices does not exist for diagrams of genera $g\geq 1$ (which also invalidates the tensor large $N=D$ limit). Indeed, for any $g\geq 1$, one can construct diagrams which are proportional to an arbitrarily high power of $D$. For example, consider the genus one graphs depicted in Fig.\ \ref{figure6}, with interaction vertices $\tr X_{\mu}X_{\nu}X_{\mu}X_{\nu}$. Note that these graphs are allowed in the Hermitian case but are forbidden in the complex case, because the self-contractions of the vertices would violate the orientation of the propagators. It is straightforward to check that the graph containing $q$ interaction vertices is proportional to $D^{1+q/2}$. There is thus no upper bound for the power of $D$. The graphs of Fig.\ \ref{figure6} can be straightforwardly generalized by using the melonic move depicted on the left of Fig.\ \ref{figure2}. In particular, we can build in this way a large class of graphs with no self-contractions on the vertices\footnote{If only graphs with self-contractions were harmful, it would be easy to get rid of them in some cases, for example by using a normal-ordered interaction at zero temperature, or dimensional regularization in massless models in space-time dimension $d\geq 2$.} that make the large $D$ limit inconsistent. It is also very easy to build similar graphs at any genus, by inserting the basic structure of the genus one graph in a larger graph.

\begin{figure}
\centering
\includegraphics[width=6in]{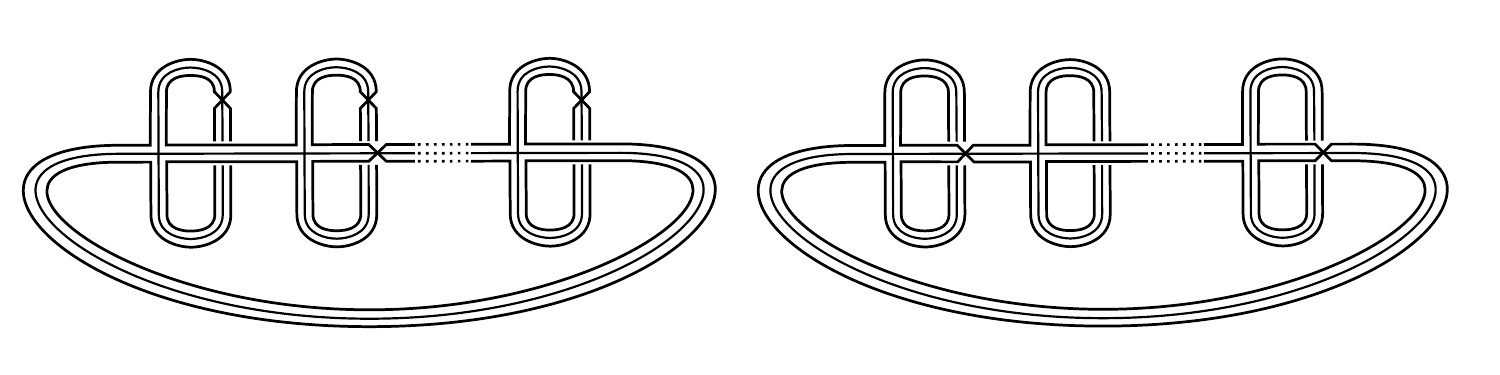}
\caption{Graphs for the real rank three tensor models studied in \cite{klebanov}. Left inset: graphs allowed in the $\text{O}(N)^{3}$ Carrozza-Tanasa model, which are at most proportional to $N^{2}$. The graph with $q$ interaction vertices must have $q\ \text{mod}\ 2$ twisted ``horizontal'' ribbons, whereas all the self-contractions come with twisted ribbons. Right inset: similar graphs allowed when the symmetry is broken down to $\text{O}(N)^{2}$ or $\text{O}(N)$ by imposing a symmetry constraint on the tensor; these graphs are proportional to an arbitrarily high power of $N$ and the large $N$ limit does not exist for these models.}
\label{figure7}
\end{figure}

Another closely related example is the $\text{O}(N)^{3}$-symmetric Carrozza-Tanasa model, or the similar $\text{O}(N)^{2}\times\text{O}(D)$-symmetric model of real matrices with an interaction term $\tr X_{\mu}X_{\nu}^{T}X_{\mu}X_{\nu}^{T}$. In the fat graph representation, the $XX$ and $XX^{T}$ propagators are twisted and untwisted ribbons respectively. The graphs of Fig.\ \ref{figure6} are thus not allowed. There are similar allowed graphs, represented in the left inset of Fig.~\ref{figure7}, but these are never greater than $N^{2}$ and are thus harmless in the large $N=D$ limit. On the other hand, if one breaks the $\text{O}(N)^{2}\times\text{O}(D)$ symmetry down to $\text{O}(N)\times\text{O}(D)$ by imposing that the matrices $X_{\mu}$ are symmetric, or down to $\text{O}(N)$ by imposing the complete symmetry between the three indices in the tensor case $N=D$, as was suggested in \cite{klebanov}, then the $XX$ and $XX^{T}$ propagators can both be either twisted or untwisted. The graphs of Fig.~\ref{figure6} are then allowed, together with the more general graphs depicted on the right inset of Fig.~\ref{figure7}. These are proportional to $N^{1+q/2}$ for any $q$, showing that the large $N$ limit of such models does not exist.

A common feature of all the ``bad'' graphs mentioned above, including their most general versions obtained by applying the melonic moves of Fig.\ \ref{figure2} on the internal edges an arbitrary number of times, is that they all contain so-called \emph{singular} edges. A singular edge is defined \cite{graphbook} to be an edge in the graph which is traversed twice by the same face. In other words, in the ribbon representation, the two borders of the ribbon associated with a singular edge belong to the same face. A simple and elegant way to eliminate all such graphs is to impose a tracelessness condition on the matrices, since singular edges are automatically associated with the contraction of two indices of the same matrix (or tensor). This yields the natural conjecture that \emph{traceless} symmetric or Hermitian models could have well-defined large $D$ limits \emph{at all genera.} Support for this conjecture is given in the recent work \cite{klebanov2}, which constructs and checks numerically a large class of graphs in the symmetrized Carrozza-Tanasa model.\footnote{We would like to thank Igor Klebanov for a fruitful exchange about this point. The harmful graphs of Fig.\ \ref{figure6} and \ref{figure7} that we communicated to the authors of \cite{klebanov2} made them realize that the consistency of the large $N$ limit required, and might be achieved by, the tracelessness condition.} It is unknown whether this conjecture will turn out to be true in full generality, for all types of interaction terms, which would suggest that a general conceptual proof, in the spirit of \cite{Fer1,FRV}, could be devised, or whether it will work only for some very specific models. For example, a detailed discussion of an interesting bipartite model can be found in \cite{guraunew}. This model is very special because the bipartite structure actually implies that the tracelessness condition is not needed.

\section{\label{CorrFunSec} On correlation functions}
\subsection{General remarks}

Based on the results derived in Section \ref{largeNDSec} for the free energy, we can analyze some properties of the large $N$ and large $D$ expansions of correlation functions in our models.

Let us consider correlation functions of general multi-trace operators. These operators may be included in the Lagrangian \eqref{lagrangian} as interaction terms $I_{\gB_a}$. They can be obtained in the usual way by taking the derivative of the free energy with respect to the associated coupling constant $\lambda_a$, taking into account the scalings in the Lagrangian \eqref{lagrangian} and in  \eqref{scaling},
\be \label{Corrderiv} N^{2-t(\gB_a)}D^{1+t(\gB_a) - c(\gB_a) + g(\gB_a)} \langle I_{\gB_a} \rangle = \frac{\partial F}{\partial \lambda_a} \, \cdotp \ee
The expansions \eqref{Fexp} and \eqref{Fhexp} for the free energy thus yield
\be \label{Corrintexp} \langle I_{\gB_a} \rangle\sim\frac{1}{N^{2-t(\gB_a)}D^{1+t(\gB_a) - c(\gB_a) + g(\gB_a)}} \sum_{h\in \mathbb N, \, \ell \in \mathbb N} N^{2-h}D^{1+\frac{h}{2}-\frac{\ell}{2}} \, . \ee

Let us consider a multiply-connected interaction term $I_{\gB}=I_{\gB_1}I_{\gB_2} \ldots I_{\gB_c}$, with $c(\gB)=c > 1$, $t(\gB_i)=t_i$, $c(\gB_i)=1$ and $g(\gB_i)=g_i$ for $i=1, \ldots, c$. By definition, we have $t(\gB)=\sum_i t_i$ and $g(\gB)=\sum_i g_i$. The expectation value $\langle I_{\gB}\rangle$ is given by \eqref{Corrintexp}. Moreover, 
\be \label{Corrintexp2}
N^{2c-t(\gB)}D^{t(\gB) + g(\gB)}\langle I_{\gB_1}\rangle \langle I_{\gB_2}\rangle \ldots \langle I_{\gB_c}\rangle = \frac{\partial F}{\partial \lambda_1}\frac{\partial F}{\partial \lambda_2} \cdots \frac{\partial F}{\partial \lambda_c}\,\cvp\ee
which is consistent with factorization at leading order,
\be\label{factorize}\langle I_{\gB}\rangle= \langle I_{\gB_1}\rangle \langle I_{\gB_2}\rangle \ldots \langle I_{\gB_c}\rangle\sim N^{t(\gB)}D^{c-t(\gB)-g(\gB)}\, .\ee
On the other hand, the connected correlation function is given by
\be \label{Corrintexp3}
N^{2c-t(\gB)}D^{t(\gB) + g(\gB)}\langle I_{\gB}\rangle_{\text c} = \frac{\partial^c F}{\partial \lambda_1 \ldots \partial \lambda_c}\, \cvp\ee
yielding at leading order
\be\label{corr222}\langle I_{\gB}\rangle_{\text c}\sim N^{2-2c+t(\gB)}D^{1-t(\gB)-g(\gB)}\, .\ee
Of course, the graphs contributing to $\langle I_{\gB}\rangle_{\text c}$ cannot be maximally disconnected and the connected expectation value is thus suppressed compared to $\langle I_{\gB}\rangle$ at leading order.

\subsection{Connected $2n$-point correlation functions} \label{conn2nCorrSec}

\begin{figure}
\centerline{\includegraphics[width=6in]{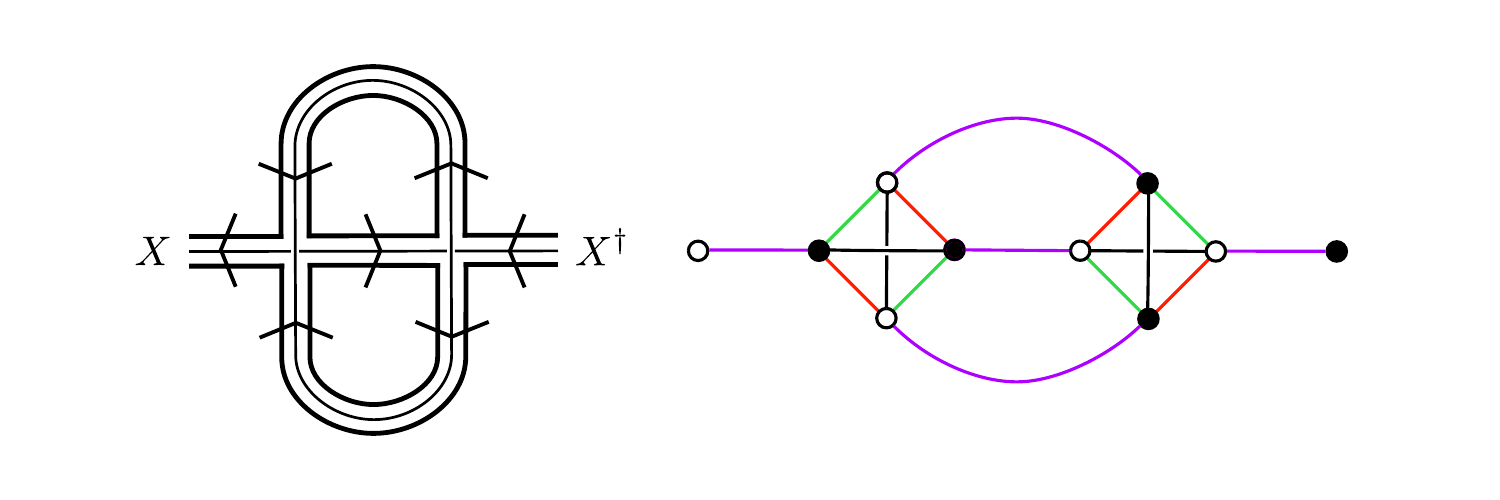}}
\caption{Example of a Feynman graph with two external legs, in the stranded and colored representations.}
\label{figureOpenGraph}
\end{figure}

In this section, we examine the large $N$ and large $D$ expansions of the connected $2n$-point correlation functions of the form
\be \label{conn2nCorr} \left\langle (X_{\mu_1})^{\alpha_1}_{\ \beta_1} (X_{\mu_2}^\dagger)^{\alpha_2}_{\ \beta_2} \ldots (X_{\mu_{2n-1}})^{\alpha_{2n-1}}_{\ \beta_{2n-1}} (X_{\mu_{2n}}^\dagger)^{\alpha_{2n}}_{\ \beta_{2n}} \right\rangle_{\text c}\, . \ee
The Feynman graphs that contribute to these correlation functions have $2n$ external legs. In the colored graph representation, the external legs correspond to ``external'' vertices of valency one, to which only a line of color 0 is attached. The external vertices are, as usual, unfilled or filled depending on whether they correspond to $X$ or $X^\dagger$. An example is depicted on Fig.\ \ref{figureOpenGraph}. We denote by $V_{\text{ext}}=2n$ the number of external vertices in the colored graph, by $V_{\text{int}}$ the number of standard internal vertices and by $V=V_{\text{ext}}+V_{\text{int}}$ the total number of vertices.

The amplitude for a Feynman diagram with $2n$ external legs is obtained from the Lagrangian \eqref{lagrangian} and the scalings \eqref{scaling} and is proportional to
\be \label{openFeynDN} N^{-p+2 v -\sum_a t(\gB_a)+ f}D^{-p+v + \sum_a (t(\gB_a)-c(\gB_a)+g(\gB_a)) + \varphi} = N^{2-2n-h^\prime}D^{1- \frac{3}{2}n+\frac{h^\prime}{2}  - \frac{\ell^\prime}{2}} \ , \ee
where we have defined the parameters $h'$ and $\ell'$ as
\begin{align}
\label{hprexp1}
h^\prime & =2+p-2v+\sum_a t(\gB_a)-f -2n \ , \\
\label{ellprexp1}
\frac{\ell^\prime}{2} & = 2+\frac{3}{2}p-2v-\frac{1}{2}\sum_a t(\gB_a) +\sum_a c(\gB_a)-\sum_a g(\gB_a)-\frac{1}{2}f-\varphi -\frac{5}{2}n \ .
\end{align}
These definitions are convenient because, as we shall prove, $h'\geq 0$ and $\ell'\geq 0$. The formulas \eqref{hprexp1} and \eqref{ellprexp1} are of course very similar to \eqref{hexp1} and \eqref{ellexp1}, with additional contributions depending on $n$.

\subsubsection{Counting the power of $N$} \label{NpowerbisSec}

We follow the same strategy as in Section \ref{NpowerSec}: we remove the $\text{O}(D)$ lines in the stranded representation and we consider the resulting matrix model fat graph. The difference is that the fat graph is now dual to a surface with boundaries because of the external insertion points. The relevant Euler's formula is
\be \label{Euleropenfatgraph} 2B^{(3)}-2g-b=-p+\tilde{v} + 2n + f = -p+\sum_a t(\gB_a) +2n + f \, , \ee
where $g$ is the genus of the fat graph, $b$ its number of boundaries and $\tilde{v}=\sum_a t(\gB_a)$ its number of internal effective single-trace vertices. Note that using the relations \eqref{fatcolid}, which are still valid here, together with
\be\label{rel34} 2 E(\gB^{(3)}) = 3 V_{\text{int}}(\gB) + V_{\text{ext}}(\gB)=3 V_{\text{int}}(\gB) + 2n\, ,\ee
where $E(\gB^{(3)})$ is the number of edges of $\gB^{(3)}$, a similar Euler's formula can be written for the colored graph,
\be\label{Euleropencolgraph}
\begin{split}
2B^{(3)}-2g(\gB^{(3)}) -b(\gB^{(3)}) & = F(\gB^{(3)}) - E(\gB^{(3)}) + V(\gB^{(3)}) \\
& = -\frac{1}{2} V_{\text{int}}(\gB) +\frac{1}{2} V_{\text{ext}}(\gB) + F_{01}(\gB) + F_{02}(\gB)+F_{12}(\gB) \\ 
& = 2B^{(3)}-2g-b \ .
\end{split}
\ee
Finally, \eqref{hprexp1} together with \eqref{Euleropenfatgraph} yields
\be \label{hprexp2} \frac{h^\prime}{2} = g + \frac{b}{2}+\sum_a \bigl(t(\gB_a)-1\bigr)-B^{(3)} +1\ , \ee
generalizing \eqref{hexp2} and showing that $h'\geq 0$ in all cases, with actually $h'\geq 1$ as soon as $n\not = 0$ since then, $b\geq 1$. As a result, we get a well-defined large $N$ expansion, the leading graphs with respect to $N$ corresponding as usual to maximally disconnected planar graphs with a single boundary component.

\subsubsection{Counting the power of $D$} \label{DpowerbisSec}

Following Section \ref{DpowerSec}, we consider $\gB^{(1)}$ and $\gB^{(2)}$, which are open three-colored graphs just like $\gB^{(3)}$, so that the corresponding Euler's formulas read 
\be\label{Euleropencolgraph2} 2B^{(i)}-2g(\gB^{(i)})-b(\gB^{(i)}) = -\frac{1}{2} V_{\rm int}(\gB) + \frac{1}{2} V_{\rm ext}(\gB) + \sum_{\substack{j<k \\ j,k\neq i}}F_{jk}(\gB) \ . \ee
On the other hand, the three-bubble $\gB^{(0)}$ remains a closed colored graph, with a standard Euler's formula. Summing these three Euler's identities and using  \eqref{fatcolid2}, we can rewrite $\ell^\prime$ in \eqref{ellprexp1} as 
\begin{multline} \label{ellprexp2}
\frac{\ell^\prime}{2} = g(\gB^{(1)})+ g(\gB^{(2)}) + \frac{1}{2} \bigl(b(\gB^{(1)})+ b(\gB^{(2)})\bigr) + \bigl({B}^{(\text{01})} - {B}^{(\text{1})} - {B}^{(\text{0})} + B\bigr)  \\
+ \bigl({B}^{(\text{02})} - {B}^{(\text{2})} - {B}^{(\text{0})} + B\bigr) + 2\Bigl(1 + \sum_a \bigl(c(\gB_a)-1\bigr) - B \Bigr)\, .
\end{multline}
As soon as $n\not = 0$, $b(\gB^{(i)})\geq1$ for $i=1,2$ and thus we get $\ell'\geq 2$. 

The above results show that the connected $2n$-point correlation functions \eqref{conn2nCorr} have well-defined large $N$ and large $D$ expansions of the form
\be \label{Correxp} \sum_{h^\prime \in \mathbb N_{\geq1}, \, \ell^\prime \in \mathbb N_{\geq2}} N^{2-2n-h^\prime}D^{1-\frac{3}{2}n+\frac{h^\prime}{2} -\frac{\ell^\prime}{2}} \, . \ee
The leading order contribution is proportional to $N^{1-2n}D^{\frac{1}{2}-\frac{3}{2}n}$, which corresponds to $h^\prime=1$ and $\ell^\prime=2$.

Let us finally note that the correlation functions of $\text{U}(N)^2$ or $\text{U}(N)^2 \times \text{O}(D)$ invariant operators can be obtained from \eqref{conn2nCorr} by contracting the free indices. This amounts to sewing together the external legs of the Feynman graphs. We let the reader rederive Eq.\ \eqref{corr222} in this way, starting from \eqref{Correxp}.

\section{\label{ModBuildSec} Model building}

The class of matrix-tensor models that can be built using the above ideas is very large and their strongly coupled physics is likely to display a wide variety of interesting new effects. On top of the SYK-like behavior \cite{Kitaev}, which is associated with a non-trivial IR limit and a macroscopic degeneracy of the ground state, it was recently discovered in \cite{AFS} that many other phenomena can occur and that the phase diagrams of the models can have a rich structure. Clearly, only the surface of this subject has been scratched and many examples, in various dimensions, remain to be studied. The aim of the present section is to provide a brief overview, emphasizing a few models that we find particularly interesting. In particular, we provide more details on some of the results for bosonic models announced in \cite{AFS}. We also explain that our new large $D$ limit, with the scaling \eqref{scaling}, is compatible with linearly realized supersymmetry. The detailed study of the physics and the phase diagrams of supersymmetric models in various dimensions is an outstanding research avenue for the future.

\subsection{\label{bossub} Unstable bosonic models}

\subsubsection{Simple models}

The simplest purely bosonic and non-trivial quantum mechanical model one can study is based on the Lagrangian
\be\label{H1bos} L = ND \tr\Bigl(\frac{1}{2}\dot X_{\mu}\dot X_{\mu}-\frac{m^{2}}{2}X_{\mu}X_{\mu}-\frac{\la^{3}}{4}\sqrt{D}\, X_{\mu}X_{\nu}X_{\mu}X_{\nu}\Bigr)\, ,\ee
where the matrices $X_{\mu}$ are Hermitian. There are obvious generalizations is any number of space-time dimension $d\leq 4$ (for $d>4$, the model is not renormalizable). This model is solvable because the leading order graphs can be fully classified following \cite{Fer1,CarrozzaTanasa}. A very similar model, which has exactly the same physics at leading order, is based on real matrices $X_{\mu ab}$ and a potential proportional to $\tr X_{\mu}X_{\nu}^{T}X_{\mu}X_{\nu}^{T}$. When $N=D$, this coincides with a special case of the Carrozza-Tanasa model \cite{CarrozzaTanasa} and is also discussed in \cite{klebanov,klebanov3}.

At leading $N\rightarrow\infty$ and $D\rightarrow\infty$ order, the solution of the model is governed by the finite temperature $T=1/\beta$ Euclidean two-point function
\be\label{Gdef1} G(t) =\frac{1}{N}\bigl\langle \tr\text{T}X_{\mu}(t)X_{\mu}\bigr\rangle_{\beta}\, ,\ee
which can be expanded in terms of Fourier-Matsubara modes as
\be\label{Fourier1} G(t) = \frac{1}{\beta}\sum_{k\in\mathbb Z}G_{k}e^{-i\nu_{k}t}\, .\ee
The Matsubara frequencies are defined by
\be\label{Matfreq} \nu_{k} = 2\pi k T\, .\ee
The structure of the dominating generalized melonic graphs implies the following  Schwinger-Dyson equations,
\begin{align} \label{SDbos1} \frac{1}{G_{k}} & = \nu_{k}^{2}+m^{2} + \Sigma_{k}\\
\label{SDbos2} \Sigma(t) &= -\la^{6}G(t)^{3}\, .
\end{align}
The self-energy $\Sigma(t)$ is expanded in Fourier series with coefficients $\Sigma_{k}$ in a way similar to \eqref{Fourier1}. At zero temperature, \eqref{Fourier1} and \eqref{SDbos1} are replaced by
\be\label{Fourier2} G(t)=\frac{1}{2\pi}\int_{-\infty}^{+\infty}\tilde G(\omega) e^{-i\omega t}\,\d\omega\ee
and
\be\label{SDbos1b} \frac{1}{\tilde G(\omega)}=\omega^{2}+m^{2}+\tilde\Sigma (\omega)\, .\ee
\subsubsection{\label{naivewrongSec}Why the naive analysis is wrong}

The equations \eqref{SDbos1} and \eqref{SDbos2} look very similar to the Schwinger-Dyson equations governing the solution of fermionic models, like the original SYK model \cite{Kitaev}. Naively, one may thus expect that the resulting physics would be very similar, but this turns out to be erroneous \cite{AFS}.

To understand in a simple way where the problem comes from, let us consider the case $m=0$ and let us try to perform the usual analysis of the IR limit of equations \eqref{SDbos1b} and \eqref{SDbos2} at zero temperature,
\be\label{SDbosIR}\frac{1}{\tilde G(\omega)}  = \tilde\Sigma (\omega)\, ,\quad
\Sigma (t) = -\la^{6}G(t)^{3}\, .\ee
Using naively the Fourier transform formula
\be\label{Fourierformula} \int_{-\infty}^{+\infty}\frac{e^{i\omega t}}{|t|^{2\Delta}}\,\d t = \frac{2\Gamma (1-2\Delta)}{|\omega|^{1-2\Delta}}\sin(\pi\Delta)\ee
and seeking power-law solutions to \eqref{SDbosIR}, $G(t)= b/|t|^{2\Delta}$ and $\Sigma(t)=b'/|t|^{2\Delta'}$, we get
\be\label{s1fake} \Delta'=1-\Delta\, , \quad \frac{1}{4b b'}=\Gamma (1-2\Delta) \Gamma (1-2\Delta') \sin(\pi\Delta)\sin(\pi\Delta')\ee 
from the first equation in \eqref{SDbosIR} and
\be\label{s2fake} \Delta' = 3\Delta\, ,\quad b'=-\la^{6}b^{3}\ee
from the second equation in \eqref{SDbosIR}. This yields
\be\label{s3fake} \Delta = \frac{1}{4}\, \cvp\quad \Delta' = \frac{3}{4}\, \cvp\quad b=\frac{1}{(4\pi\la^{6})^{1/4}}\ee
and produce the following ``solution,''
\be\label{fakes}
\begin{aligned}& G(t) = \frac{1}{(4\pi\la^{6})^{1/4}}\frac{1}{\sqrt{|t|}}\,\cvp\quad \tilde G(\omega) = \Bigl(\frac{\pi}{\la^{6}}\Bigr)^{\frac{1}{4}}\frac{1}{\sqrt{|\omega|}}\, , \\&
\Sigma(t) =-\la^{6}\frac{1}{(4\pi\la^{6})^{3/4}}  \frac{1}{|t|^{\frac{3}{2}}}\, \cvp\quad\tilde\Sigma(\omega) = \Bigl(\frac{\la^{6}}{\pi}\Bigr)^{\frac{1}{4}}\sqrt{|\omega|}\, .
\end{aligned}
\ee
However, this result is inconsistent. On the one hand, it predicts $\tilde G(\omega)>0$ and $\tilde\Sigma(\omega)>0$, whereas the Fourier transform of the second equation in \eqref{SDbosIR},
\be\label{SDbosFourier} \tilde\Sigma(\omega) =-\frac{\la^{6}}{4\pi^{2}}\int \tilde G(\omega_{1})\tilde G(\omega_{2})\tilde G(\omega-\omega_{1}-\omega_{2})\, \d\omega_{1}\d\omega_{2}\, ,\ee
clearly shows that $\tilde G(\omega)>0$ implies that $\tilde\Sigma(\omega)<0$. There is no way \eqref{fakes} could be a meaningful solution of \eqref{SDbosIR}.

The mistake comes from the fact that the above reasoning, albeit standard in the literature, involves formal manipulations of divergent integrals. The Fourier transform formula \eqref{Fourierformula} makes sense only if $\Delta>0$ to avoid IR divergences and $\Delta<1/2$ to avoid UV divergences. The analysis 
however assumes that one can use \eqref{Fourierformula} to compute the Fourier transforms of both $G$ and $\Sigma$; this is clearly incompatible with the first equation in \eqref{s1fake}, since $\Delta<1/2$ implies that $\Delta'=1-\Delta>1/2$.\footnote{This problem does not occur, e.g., in the usual SYK model, because $G$ and $\Sigma$ are then odd functions of time and the Fourier transforms involve the integrals of $|t|^{-2\Delta}\sin(\omega t)$ and $|t|^{-2\Delta'}\sin(\omega t)$, which are not UV divergent.}

One may think that the difficulty comes from the use of the IR limit of the exact Schwinger-Dyson equation \eqref{SDbos1b}, but this is not the case. It is easy to see that the full set of equations \eqref{SDbos1} cannot have consistent solutions when $m\rightarrow 0$. The argument uses unitarity. It is straightforward to show, using the spectral decomposition of the two-point function \eqref{Gdef1}, that the Fourier coefficients $G_{k}$ must be real and strictly positive. But when $m\rightarrow 0$, \eqref{SDbos1} implies that $\Sigma_{0}>0$ too. One then obtains a contradiction with \eqref{SDbos2}, which shows that $\Sigma_{0}$ is a sum of strictly negative terms.

Let us note that the above conclusions are not restricted to the case of quantum mechanics. The discussion can indeed be straightforwardly generalized to the case of higher dimensional field theories, of the type considered for example in \cite{klebanov3}.

\subsubsection{The physics}

The physics associated with the above phenomenon is explained in \cite{AFS} and will be discussed further in \cite{AFS2}. Let us simply briefly recall the main points here. The difference between bosonic models like \eqref{H1bos} and SYK-like fermionic models is twofold. First, unlike in the fermionic cases, the large temperature limit of bosonic models is not weakly coupled in general. For instance, the model \eqref{H1bos} at $m=0$ is always strongly coupled, even when the dimensionless ``coupling'' $\beta\la$ is very small. An SYK-like high temperature pertubation theory thus simply does not exist for bosonic models. Second, models like \eqref{H1bos} are unstable. The leading large $N$ and large $D$ limits still make sense, but only as long as the effective dimensionless coupling is not too strong. The particular point $T=0$, $m\rightarrow 0$ in parameter space discussed in \S\ref{naivewrongSec} belongs to a larger strongly coupled region where the equations \eqref{SDbos1} and \eqref{SDbos2} do not have a solution, see Fig.\ 5 in ref.\ \cite{AFS}. 

\subsection{Stable bosonic models}

One can easily build stable purely bosonic models, for which the large $N$ and large $D$ limits are still dominated by generalized melonic diagrams and can thus be exactly solved. These models are interesting for several reasons. For example, one would like to investigate whether the absence of a non-trivial IR limit for the model \eqref{H1bos} is due to its instability or to other qualitative differences with the fermionic models, such as the absence of a high temperature perturbation theory \`a la SYK.

A simple Lagrangian with Hermitian matrices and a stable potential is
\be\label{H2bos} L = ND \tr\Bigl(\frac{1}{2}\dot X_{\mu}\dot X_{\mu}-\frac{m^{2}}{2}X_{\mu}X_{\mu}-\frac{\la^{4}}{2} D \, X_{\rho}X_{\mu}X_{\rho}X_{\sigma}X_{\mu}X_{\sigma}\Bigr)\, .\ee
The scaling with $D$ of the interaction term is according to \eqref{scaling}. The potential is manifestly positive, since it can be rewritten as $(1/2) \tr A_{\mu}A_{\mu}$ where
\be\label{Adef} A_{\mu}=\sqrt{D}\la^{2}X_{\rho}X_{\mu}X_{\rho}\ee
is Hermitian. The leading Feynman graphs can be easily classified by introducing an auxiliary field $F_{\mu}$ and noting that \eqref{H2bos} is equivalent to
\be\label{H3bbis} L=ND \tr\Bigl(\frac{1}{2}\dot X_{\mu}\dot X_{\mu}-\frac{m^{2}}{2}X_{\mu}X_{\mu}+\frac{1}{2}F_{\mu}F_{\mu}-\la^{2}\sqrt{D}\, F_{\mu}X_{\nu}X_{\mu}X_{\nu}\Bigr)\, .\ee
The interaction $\tr F_{\mu}X_{\nu}X_{\mu}X_{\nu}$ is of the same general form as in \eqref{H1bos}. Moreover, the scaling with $D$ that results from going from \eqref{H2bos} to \eqref{H3bbis} is nicely consistent with \eqref{scaling}. This is a general fact whose generalization has nice consequences in supersymmetric models, as we will explain below. We can thus use the results of \cite{Fer1,CarrozzaTanasa} to get the leading graphs. Introducing the Euclidean two-point functions $G_{X}(t)$, $G_{F}(t)$ and the associated self-energies and  Fourier transforms, the Schwinger-Dyson equations read, in obvious notations,
\begin{align}\label{sdstab1}& \frac{1}{G_{X,k}}=\nu_{k}^{2}+m^{2}+\Sigma_{X,k}\, ,\quad
\frac{1}{G_{F,k}}=-1+\Sigma_{F,k}\, , \\\label{sdstab2} &\Sigma_{X}(t)=-3\la^{4}G_{X}(t)^{2}G_{F}(t)\, ,\quad \Sigma_{F}(t)=-\la^{4}G_{X}(t)^{3}\, .
\end{align}
Note that the Fourier coefficients $G_{A,k}$ for the two-point function of the operator $A_\mu$ defined in \eqref{Adef} are simply given by $G_{A,k}=G_{F,k}+1$.

We can proceed to analyze \eqref{sdstab1} and \eqref{sdstab2} in the usual way. A naive solution of the IR limit of the equations with non-trivial scaling dimensions $1/6$ and $1/2$ associated with the two-point functions $G_{X}$ and $G_{A}$ can be found straightforwardly. However, a careful analysis shows that this solution actually does not exist, that is to say, it is \emph{not} the IR limit of a solution to the full set of equations \eqref{sdstab1} and \eqref{sdstab2} \cite{AFS, AFS2}. This result is surprising. At least when $m\rightarrow 0$, it would have been natural to guess that the model \eqref{H2bos} could develop a non-trivial IR behavior, with a non-zero zero temperature entropy, etc.; pretty much in the same way as the fermionic SYK-like models do. The fact that it does not shows that there is an important qualitative difference between purely fermionic and purely bosonic models, \emph{even when the bosonic models are stable and dominated by the same type of generalized melonic diagrams as the fermionic models.} To the best of our knowledge, no purely bosonic model with SYK behaviour has been found up to now.

\subsection{Supersymmetric models}

Supersymmetric theories are extremely natural to look at, in particular if one wishes to devise models with a gravitational dual. Up to now, most studies have focused on models with quenched disorder and/or a non-linear realization of supersymmetry where all fundamental degrees of freedom are fermions, see e.g.\ \cite{SYKsusy}. Here we point out that our matrix-tensor models have standard linearly realized supersymmetric versions with two or four supercharges.\footnote{See also \cite{klebanov} for linearly realized supersymmetrization of 
an SYK-like tensor model.}

There is one possible obstruction to build models with linearly realized supersymmetry: supersymmetry relates several interaction terms together and this is not obviously consistent with the large $D$ scaling \eqref{scaling}. Our simple goal in the present subsection is to display explicitly how supersymmetry acts on the bubble representing the interaction terms and check that this action is consistent with the scaling \eqref{scaling}. Note that another way to understand the consistency of sypersymmetry with \eqref{scaling} is to use a supergraph formulation of the Feynman rules. 

We focus on $\mathcal N=2$ supersymmetric matrix quantum mechanics for concreteness. The case $\mathcal N=4$ and higher space-time dimensions are very similar. The $\mathcal N=2$ models contain traceless Hermitian bosonic matrices $X_{\mu}$ and complex fermionic matrices $\psi_{\mu}$ transforming 
in the adjoint representation of $\uN$. The real superpotential can be written in parallel with the interaction terms in the Lagrangian, see Eqs.\ \eqref{lagrangian} and \eqref{interterm}, in the form
\be\label{Wexp} W(X)=\sum_{a}N^{1-t(\gB_{a})}\tau_{a}\mathscr {I}_{\gB_{a}}(X)\, .\ee
The bubbles $\gB_{a}$ encode the term $\mathscr {I}_{\gB_{a}}(X)$ in the usual way. The superpotential yields two types of interaction terms in the Lagrangian.

The terms coupling the fermions and the bosons read
\be\label{BosFer}L_{1} =-ND\, \bar{\psi}_{\mu}{}^\alpha{}_{\beta}\frac{\partial^2 W}{\partial X_{\mu}{}^\alpha{}_\beta \partial X_{\nu}{}^\gamma{}_\delta} \psi_{\nu}{}^\gamma{}_\delta=
-ND\sum_{a}N^{1-t(\gB_{a})}\tau_{a} I_{\gB_{a}}(X,\psi,\bar\psi)\, ,\ee
where
\be\label{BosFer2} I_{\gB_{a}}(X,\psi,\bar\psi) = \bar{\psi}_{\mu}{}^\alpha{}_{\beta}\frac{\partial^2 \mathscr {I}_{\gB_{a}}}{\partial X_{\mu}{}^\alpha{}_\beta \partial X_{\nu}{}^\gamma{}_\delta} \psi_{\nu}{}^\gamma{}_\delta\, .\ee
Each term in $I_{\gB_{a}}$ is obtained from $\mathscr {I}_{\gB_{a}}$ by substituting two matrices $X_{\mu}$ and $X_{\nu}$ by $\psi_{\mu}$ and $\bar\psi_{\nu}$. They are thus all labelled by the same bubble $\gB_{a}$ and in particular, the coupling constants $\tau_{a}$ must scale as in \eqref{scaling} at large $N$ and large $D$.

\begin{figure}
\centering
\includegraphics[width=6in]{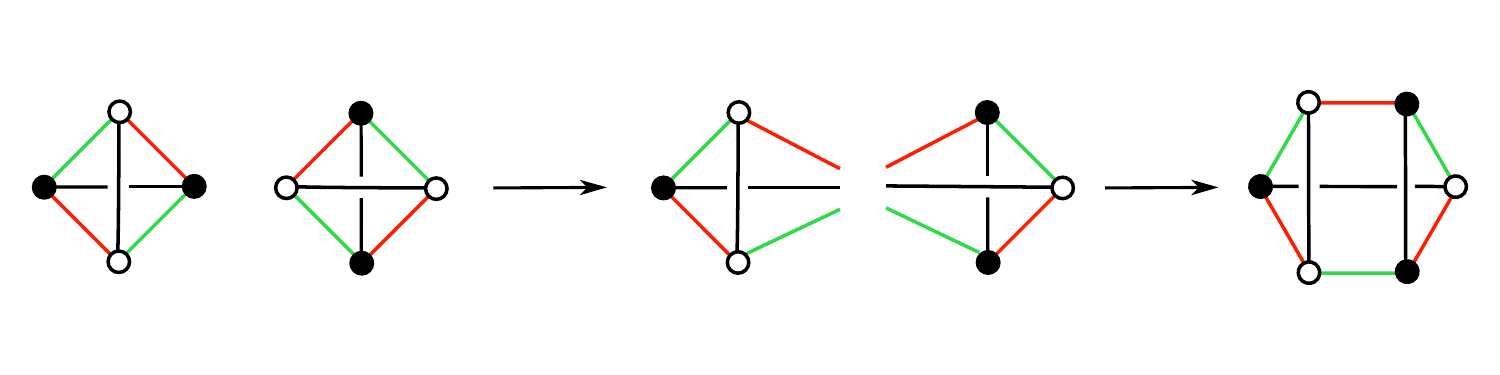}
\caption{Construction of the bubble associated with the potential term \eqref{potsusy} (right inset) from the bubble associated with the superpotential \eqref{Wexp} (left inset) in the case $W(X)=\tau \tr X_{\mu}X_{\nu}X_{\mu}X_{\nu}$, as described below Eq.\ \eqref{pot2}.
\label{figcontraction}}
\end{figure}

The potential term contributes as
\be\label{potsusy} L_{2}=-\frac{1}{2} ND  \frac{\partial W}{\partial X_{\mu}{}^\alpha{}_\beta} \frac{\partial W}{\partial X_{\mu}{}^\beta{}_\alpha}
= -\frac{1}{2} ND\sum_{a,b}N^{2-t(\gB_{a})-t(\gB_{b})}\tau_{ab}I_{ab}(X)\, ,\ee
where
\be\label{couplingsusy} \tau_{ab}=\tau_{a}\tau_{b}\ee
and
\be\label{pot2} I_{ab}(X)=
\frac{\partial \mathscr{I}_{\gB_{a}}}{\partial X_{\mu}{}^\alpha{}_\beta} \frac{\partial \mathscr{I}_{\gB_{b}}}{\partial X_{\mu}{}^\beta{}_\alpha}\,\cdotp\ee
The interaction terms appearing in $I_{ab}$ are described by many different bubbles, which we denote collectively by $\gB_{ab}$, obtained from $\gB_{a}$ and $\gB_{b}$ in the following way: we remove one vertex from $\gB_{a}$ and one vertex from $\gB_{b}$ and then join together the edges that were attached to these two vertices in a way consistent with the coloring. In the tensor model literature, this operation is called a ``one-dipole contraction.'' An example is depicted in Fig.\ \ref{figcontraction}. Since one connected component of $\gB_{a}$ is connected to one connected component of $\gB_{b}$ under the contraction, we have
\be\label{cont1} c(\gB_{ab}) = c(\gB_{a})+c(\gB_{b}) -1\, .\ee
Moreover, the vertices we remove belong to $(12)$-faces in $\gB_{a}$ and $\gB_{b}$ and under the contraction these two faces merge into one, which yields
\be\label{cont2} t(\gB_{ab}) = t(\gB_{a})+t(\gB_{b}) -1\, .\ee
By considering similarly the $(13)$- and $(23)$-faces passing through the vertices that are removed, we get
\be\label{cont3} F_{13}(\gB_{ab}) = F_{13}(\gB_{a})+F_{13}(\gB_{b}) -1\, ,
\quad F_{23}(\gB_{ab}) = F_{23}(\gB_{a})+F_{23}(\gB_{b}) -1\, .\ee
Using \eqref{cont1}, \eqref{cont2}, \eqref{cont3} together with 
\be\label{cont4} V(\gB_{ab})=V(\gB_{a})+ V(\gB_{b})-2\, ,\ee
we then obtain
\be\label{cont5} g(\gB_{ab})=g(\gB_{a})+ g(\gB_{b})\, .\ee
According to the general formulas \eqref{lagrangian} and \eqref{scaling}, the overall powers of $N$ and $D$ in front of $I_{ab}(X)$ in \eqref{potsusy} must thus be
\begin{multline}\label{overallND} ND\times N^{1-t(\gB_{ab})}\times D^{t(\gB_{ab})-c(\gB_{ab})+g(\gB_{ab})}=\\ ND\times N^{2-t(\gB_{a})-t(\gB_{b})}\times D^{t(\gB_{a})+t(\gB_{b})-c(\gB_{a})-c(\gB_{b})+g(\gB_{a})+g(\gB_{b})}\, .
\end{multline}
Comparing with \eqref{potsusy}, we see that the coupling $\tau_{ab}$ must scale as
\be\label{overallDtab} D^{t(\gB_{a})-c(\gB_{a})+g(\gB_{a})}\times D^{t(\gB_{b})-c(\gB_{b})+g(\gB_{b})}\, .\ee
This is precisely matching the scaling implied by the supersymmetric relation \eqref{couplingsusy} between couplings and by \eqref{scaling}, as was to be shown.

\section*{Acknowledgements}

We would like to thank Igor Klebanov for discussions. This work is supported in part by the Belgian Fonds National de la Recherche
Scientifique FNRS (convention IISN 4.4503.15) and the F\'ed\'eration Wallonie-Bruxelles (Advanced ARC project ``Holography, Gauge Theories and Quantum Gravity''). G.~V.~ is a Research Fellow at the Belgian F.R.S.-FNRS. T.~A.~ would like to thank Centro de Ciencias de Benasque Pedro Pascual for hospitality during his stay for the workshop ``Gravity - New perspectives from strings and higher dimensions''. The work of P.G. is partially supported by the Compagnia di San Paolo contract ``MAST:Modern Applications of String Theory'' TO-Call3-2012-0088.

\end{document}